\documentclass[12pt,preprint]{aastex}

\shorttitle{Approaching the Hydrogen Burning Limit in M4}
\shortauthors{Bedin et al.}

\begin{document}

\def\hst{{\sl HST}}


\title{COLOR--MAGNITUDE DIAGRAM AND LUMINOSITY FUNCTION OF M4 NEAR 
THE HYDROGEN-BURNING LIMIT\footnote{Based on observations with the 
NASA/ESA {\it Hubble Space Telescope}, obtained at the Space Telescope Science Institute,
which is operated by AURA, Inc., under NASA contract NAS 5-26555.}}


\author{Luigi R.\ Bedin\footnote{Visiting Student at U.\ C.\ Berkeley.}}
\affil{Dipartimento di Astronomia, Universit\`a
di Padova, Vicolo dell'Osservatorio 2,\\
I-35122 Padova, Italy; bedin@pd.astro.it}

\medskip
\author{Jay Anderson and Ivan R.\ King\affil{Astronomy Department,
University of California, Berkeley, CA 94720-3411;\\ 
jay@cusp.berkeley.edu, king@glob.berkeley.edu}}

\medskip
\author{and \ \vskip -12pt}

\author{Giampaolo Piotto\affil{Dipartimento di Astronomia, Universit\`a
di Padova, Vicolo dell'Osservatorio 2,\\ 
I-35122 Padova, Italy; piotto@pd.astro.it}}

 

\begin{abstract}
A proper-motion separation of M4 members from field stars, using deep
\hst\ observations separated by a time base-line of 5 years, allows us
to study a pure sample of cluster main-sequence stars almost to the
minimum mass for hydrogen burning.  High-precision photometry shows how
badly current theoretical models fail to reproduce the color--magnitude
diagram 
of low-mass stars of moderate metallicity ([M/H] $\simeq-$1).  
This inability of theory to reproduce the luminosity--radius relation casts
doubt on the theoretical mass--luminosity relation, which is needed to
convert the observed luminosity function (LF) into a mass function
(MF), as well as to convert our locally determined LF
into a global MF.  To the extent that we trust theoretical
M--L relations for such
transformations, we obtain a flat MF from the LF, and some indication
that theoretical masses might be too low at a given luminosity, near
the H-burning limit.

\end{abstract}




\section{INTRODUCTION}
\label{intro}

There is a minimum mass below which a contracting protostar cannot
ignite thermonuclear burning of hydrogen. Close to that H-burning limit
(HBL), which separates main sequence (MS) stars from brown dwarfs, old
stars show a huge difference of luminosity for a small difference in
mass.  This effect results in a plunge of the luminosity function (LF)
toward zero, for stars with masses just above this limit.

The best place to observe the properties of stars approaching the HBL is
in nearby Galactic globular clusters (GGCs). Not only are their stars
homogeneous in age, distance, and chemical composition, but at the
typical GGC age of 10 Gyr or more, the stars with masses below the HBL will
have faded by several magnitudes relative to those above the HBL, thus
creating a virtual cutoff in the LF.

We have already presented a preliminary study of faint MS stars in NGC
6397, the cluster of smallest apparent distance modulus (King et al.\
1998).  
Study  of NGC 6397 is continuing---higher-accuracy remeasurement
of the existing field, and observations of three more fields.  In the
present paper we give preliminary results for a second cluster, M4 (NGC
6121), whose higher metallicity introduces a new parameter into the
confrontation of theory with observation.  M4 is geometrically the
closest globular cluster to the Sun but is more obscured than NGC 6397,
leaving it as the cluster of second smallest apparent distance modulus.
Like NGC 6397, M4 is in a rich field not far from the Galactic center
($l=351^\circ,$ $b=+16^\circ$), and here too its faint stars would be
hopelessly lost among field stars were it not for our ability to use
proper motions to distinguish the cluster members.  
In fact, the large
proper motion of M4 (Cudworth \& Rees 1990) makes it an unusually good
candidate for such a study.

\section{OBSERVATIONS AND MEASUREMENTS}
\label{observation}

Three fields in M4 were observed by another group in March--April 1995
(GO-5461) with the WFPC2 camera of the Hubble Space Telescope
(\hst). Their photometric data are given by Ibata et al.\ (1999), with
color--magnitude diagrams (CMDs); and the
white dwarfs are discussed by Richer et al.\ (1995, 1997). 
Here, we supplement their work
with a study of the lower MS, enabled by
re-observations of their field in April 2000 (GO-8153).
The 5-year base-line gives cluster stars a mean displacement
of $\sim85$ mas (0.85 WF pixel) from field stars, an amount so easily
measured as to effect a near-perfect separation.

In this work we shall present results from only the outermost of
our fields. The reasons are that (1) a single field is sufficient to
lead to strong and interesting conclusions, (2) this field goes the
deepest and is the least crowded of the three, 
allowing us to present the deepest GGC LF ever obtained,
and (3) preliminary
dynamical models suggest that this field is representative of the global
mass function (MF).

Our field is about 6 core radii ($r_{\rm c}\simeq50^{\prime\prime}$) from
the cluster center.  
The first epoch consists of 15$\times$2100s in F555W and 9$\times$800s
in F814W; the second epoch has 3$\times$600s plus 5$\times$700s F814W
exposures.  All sets of images are well dithered, following the recipe
in Anderson \& King (2000).

We carried out the astrometry/photometry, for each filter and each
epoch, with algorithms based on the point-spread-function (PSF)
fitting procedure described by Anderson \& King (2000).  The essence of
the method is to determine a finely sampled PSF of high accuracy, from
images at several dither offsets.  Fitting of this PSF to individual
star images gives a positional accuracy of $\sim0.02$ pixel, without any
systematic error from the location of the star with respect to pixel
boundaries.

Stars were selected by tabulating all local peaks in our images,
establishing transformations to a common coordinate frame, and seeing in
how many images a putative star showed a peak within one pixel of its
mean position.  The optimum balance between losing good stars and
accepting false peaks turned out to fall at 9 detections out of 15
images (F555W) and 6 out of 8 or 9 (2 epochs of F814W).

Since we hold to the virtue of keeping our photometry in the
instrumental pass-bands, the only calibration question is zero points.
For these we chose the ``flight'' photometry system as defined by
Holtzman et al.\ (1995).  For F814W, which was observed at both epochs,
our magnitude for each star is the weighted mean of the two epochs.

For convenience we will refer to F814W and F555W as $I$ and $V$,
respectively, although it should be clearly understood that all
photometry presented here is in the instrumental pass-bands rather than
in any conversion to a standard system.

We discarded stars brighter than $I_{814}\simeq 18.3$, at which magnitude
even the PC starts to saturate (WF saturation starts at $I_{814}\simeq
19.2$).

\section{PROPER MOTIONS AND THE COLOR--MAGNITUDE DIAGRAM}
\label{propCMD}

The proper-motion separation is shown in Figure 1.  Since all
measurements were made with respect to reference stars that are cluster
members, the zero point of motion is the centroid motion of the cluster.
The left side of the figure shows all stars that were detected both in
$V$ and in $I$.  

Since the cluster stars show such a narrow range of proper motions, we
arbitrarily set a membership criterion of 0.3 pixel from the mean.  The
middle and right-hand parts of Fig.\ 1 show this separation.  It
is clear from the distributions of points in the upper diagrams that we
have lost very few true members, and that perhaps two or three field
stars have motions that accidentally throw them into the list of
members.
Contrary to our previous practice in NGC 6397, where we insisted that a
star be found both in $I$ and in $V$ (Cool, Piotto, \& King 1996), in M4
we delve more deeply, by including stars that were found in $I$ but not
in $V$.  In the middle and right sections of Fig.\ 1 they are plotted
arbitrarily at $V_{555}-I_{814}=4$; in the upper plots they are also
shown as open circles.  These stars include the faintest stars that we
detect on the MS of the cluster.  They are
crucial in showing that we have come close to the 
HBL, because their distribution in the center and right-hand CMDs makes
it clear that cluster stars peter out at a magnitude where field stars
are still numerous.  Without these stars the left and middle CMDs would
imply that the lower limit of the MS is set as much by the $V$
detection limit as by any true drop in the cluster LF.

The $I$-only list includes, surprisingly at first, half a dozen stars
brighter than $I_{814}=22$.  These turn out, however, to be neighbors of
bright cluster giants, which happen to interfere more in $V$ than in
$I$.

The numbers at the right of the CMDs are the number of stars, in
half-magnitude bins, found in both $V$ and $I$ (excluding white dwarfs),
followed by the number found in $I$ only.  Each set of numbers is
printed at the ordinate value that bisects its magnitude bin.

As a check on the reliability of our photometry, we have compared our
measurements with those of the original investigators of the
first-epoch material (Ibata et al.\ 1999, Table 5).  The agreement is
excellent, with some indication, however, that our sequences are
narrower.  We will give details in a paper on our methods, which is in
preparation. 

\section{COMPARISON WITH THEORETICAL ISOCHRONES}
\label{comp}

In Figure 2 is shown the comparison of our colors and magnitudes with
those predicted by Baraffe et al.\ (1997),  Cassisi et al.\ (2000),  and
by Montalban, D'Antona, and Mazzitelli (2000) for these same \hst\
filters.  The agreement is clearly quite poor. 
The $V_{555}-I_{814}$ colors of the ridge line of the CMD, in steps of
0.5 magnitudes, starting from $I_{814}=18.0$, are: 1.43, 1.63, 1.85, 2.02,
2.15, 2.28, 2.38, 2.51, 2.65, 2.82, 3.08, 3.35.

In Fig.\ 2 we adopted the reddening and the distance modulus obtained by
Richer et al.\ (1997) and Ibata et al.\ (1999), which implies (for a K5 star,
Holtzman et al.\ 1995) $E(V_{555}-I_{814})=0.45$ and $(m-M)_{I_{814}}\simeq12.0$. 
A different choice of reddening and distance modulus would not
significantly improve the fit.

As shown by Montalban et al.\ (2000, but see also discussion in Cassisi et al.\ 2000),
we believe that the bad fit is related to the higher
metallicity of M4, which pushes the theory into a realm that it is not
yet able to handle.  Whether the lack is in atmospheric opacities or
in the internal equation of state, it is plausible that the theory,
while adequately fitting the lower main sequence of the
low-metallicity cluster NGC 6397 (King \&
Anderson 2001), breaks down at the nearly-10-fold-higher metallicity
of M4.

\section{THE LUMINOSITY FUNCTION}
\label{lftxt}

Our next task is to derive a luminosity function for the faint main
sequence of M4.  First, can we be sure that the faint $I$-only stars are
really MS rather than white dwarfs?  To answer this question we examined
``super-images'' that we had built up in each color, at the first epoch,
by combining all the images in that color into a single sub-sampled
image.  
Super-images have twice the original resolution. They were
obtained by an iterative technique completely analogous to
the way we construct the effective PSF (Anderson and King 2000).
Although the super-images are unsuitable for photometry or astrometry,
they allow ready examination for possible faint stars.  It turned out
that each of the 13 faint $I$-only cluster members was weakly visible
in the $V$ super-image.  These stars must all be MS rather than white
dwarfs; if any of them had the $V$ magnitude that would correspond to
a WD color, it would be more than 10 times as bright as the weak peak
seen in the super-image.

The derivation of a luminosity function requires more than just counting
stars in magnitude bins; the most important step is the determination of
completeness as a function of magnitude.  Toward this end, following
Piotto \& Zoccali (1999), we added to each of our actual $I$ images
artificial stars, each of which was a replica of the PSF with 
Poisson noise added.  The artificial stars were
geometrically spaced in such a way that they cannot interfere with the
detection of each other.  Separate tests added the artificial stars
repeatedly to the same original image at random positionings of their
grid, until there was a large enough number to give good completeness
statistics:\ for each chip, about 2500 stars at random magnitudes from
19 to 23.8 and about 6500 more from magnitude 23.8 to 25.5.  
Because each exposure was at a different dither position, we took care
to place each artificial star at the proper position in
each image.
The artificial stars were then subjected to the same finding criteria as
the real stars in the image, and each was noted as found or not found.

A curve of completeness against magnitude was derived for each chip, and
each star was assigned a completeness value $c$ by interpolation in a
smoothed version of the completeness curve for its chip.  In each
half-magnitude bin of each chip, the effective number of stars was
calculated as $\sum 1/c_i$, with a corresponding uncertainty
$\sigma=\sqrt{\sum 1/c_i^2}$.

The PC has relatively few stars, and none near the faint end; hence we
have made our LF by summing the numbers for the three WF chips only.
It is shown in Figure 3, along with the mean completeness
curve of the three chips.  The LF points of course include allowance
for completeness.  Note that the average completeness in the last two
bins is 69\% and 31\%, respectively.  

The zero count in the last bin
is striking.  If this bin stays empty in a later LF based on a larger
sample of stars, it will point strongly to the location of the HBL.
We will discuss this region in detail in a future paper that will
exploit all three M4 fields.

\section{THE MASS FUNCTION}
\label{mf} 
To derive a mass function we need a mass--luminosity relation (MLR).
Any such relation must be suspect, since we have seen how the
theoretical stellar models fail to produce the correct relation between
magnitude and color.  We will, however, work with what is available, for
whatever it may be worth.  Available to us are absolute magnitudes in
the F814W band for stars of various masses, as predicted by Baraffe et
al.\ (1997), by Cassisi et al.\ (2000), and by Montalban et al.\
(2000). 

Our procedure was as follows: We adopted the distance modulus
$I_{814}-M_{814}=12.0$ that we used in Sect.\ \ref{comp}---although we
will examine the effect of changing this number.  For a given MLR we
then transformed the LF of Figure 3 into a MF.

Figure 4 shows the MFs.  We omitted the last bin (with zero
counts), already discussed in the previous section.  In each case the points
are consistent with a flat mass function.  The numbers at the two
points of lowest mass seem low, however, as if the masses assigned to
these faintest stars ought to be somewhat higher.

How firm is this result?  Statistically it is of course very weak, and
we look forward very much to our future fuller discussion, with several
times as many stars.  Considerations of distance modulus or cluster
modeling do not shake it, however.  We repeated the calculation with
values of 11.8 and 12.2 for the distance modulus, and the results differ
so little that they are not worth exhibiting here.  As for dynamics,
preliminary fitting of multi-mass King models
shows that this field is at a radius where,
coincidentally, the correction from local to global mass function is
rather small.

\section{CONCLUSIONS}
\label{conc} 

Using the powerful tool of proper-motion separation of cluster from
field, we have been able to follow the main sequence of M4 down to the
neighborhood of the hydrogen-burning limit.  

Even with this single small field, it is clear that existing theoretical
stellar models are unable to predict correct colors at the fainter
luminosities, at the metallicity of M4.
In spite of the discrepancy in the CMD, already evident at $\sim 0.5
m_\odot$, the MLR seems to behave properly down to $\sim0.13 m_\odot$. 
Below this, it seems to be less steep than needed, if we are to avoid a sharp 
bend in the last points of the MF, corresponding to a mass interval of 
only $\sim0.03m_\odot$.

Our sample is very small for checking mass--luminosity relations, 
but the location at which the LF becomes statistically
indistinguishable from zero suggests that the steep plunge in the LF
that heralds the H-burning limit might occur at a slightly higher mass
than theory has suggested.

Continuing study of M4, with the addition of two richer fields, will
hopefully strengthen these conclusions.

\acknowledgments We are grateful to Francesca D'Antona for advice
regarding the status of the theory, and also to her, to Santi Cassisi,
and to Gilles Chabrier and Isabelle Baraffe for sending details of
theoretical models as well as the \hst\ magnitudes that they predict.
This research was supported by STScI Grant GO-8153.  GP and LRB
recognize partial support by the MURST and by the ASI.  LRB is
grateful to the CNAA and to Fondazione ing.\ Aldo Gini for grants which
have supported his stay at U.\ C.\ Berkeley.


\clearpage

\begin{figure}
\plotone{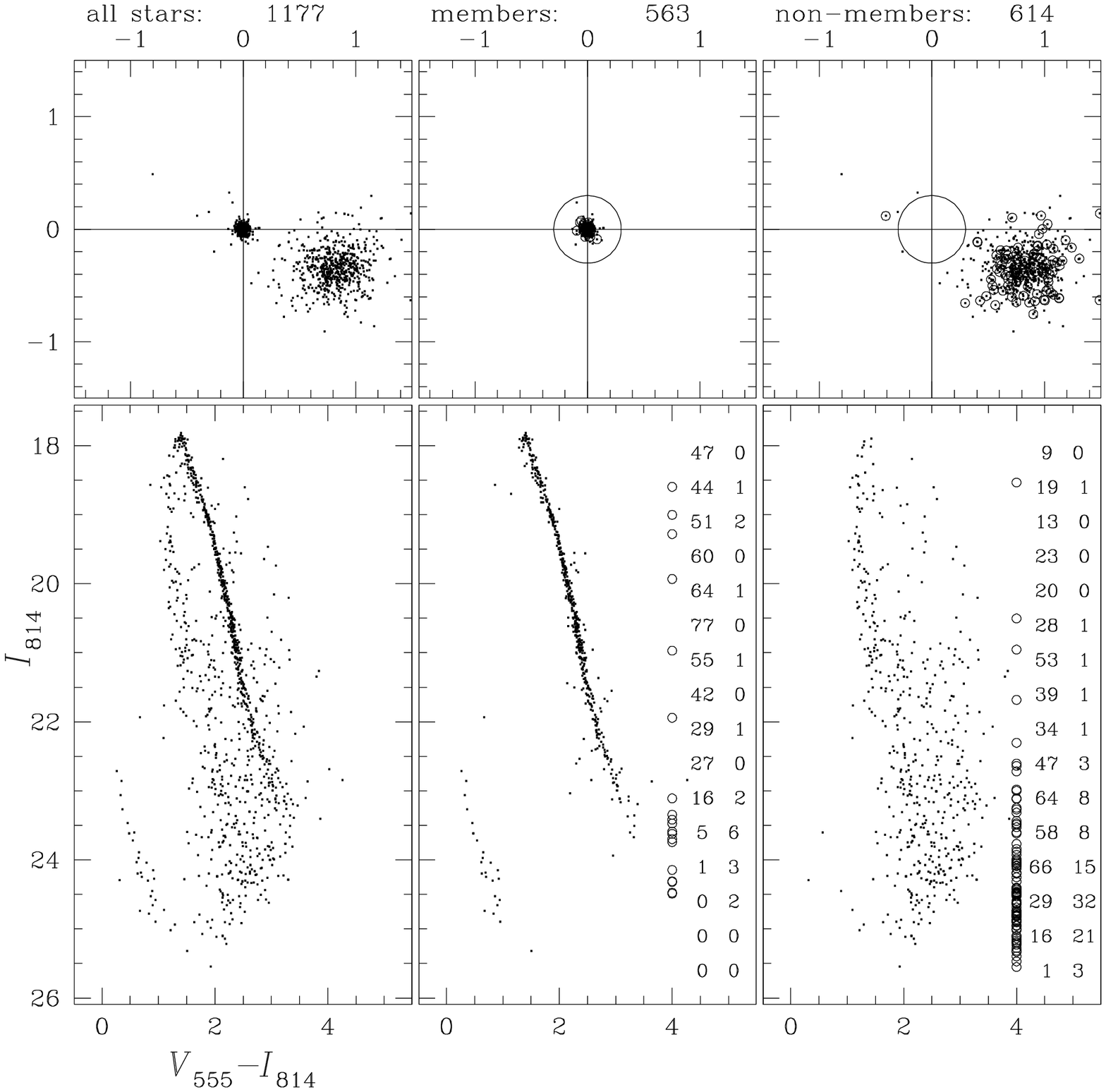}
\figcaption{Above, 5-year displacements in the $I$ images (unit = 1 WF
pixel); below, color--magnitude diagrams.  Left, all stars detected in
both $V$ and $I$; center and right, separation by the proper-motion
criterion shown.  At center and right, stars detected in $I$ but not in
$V$ are shown as open circles, plotted at an arbitrary red color.}
\label{all}
\end{figure}
\clearpage 

\begin{figure}
\plotone{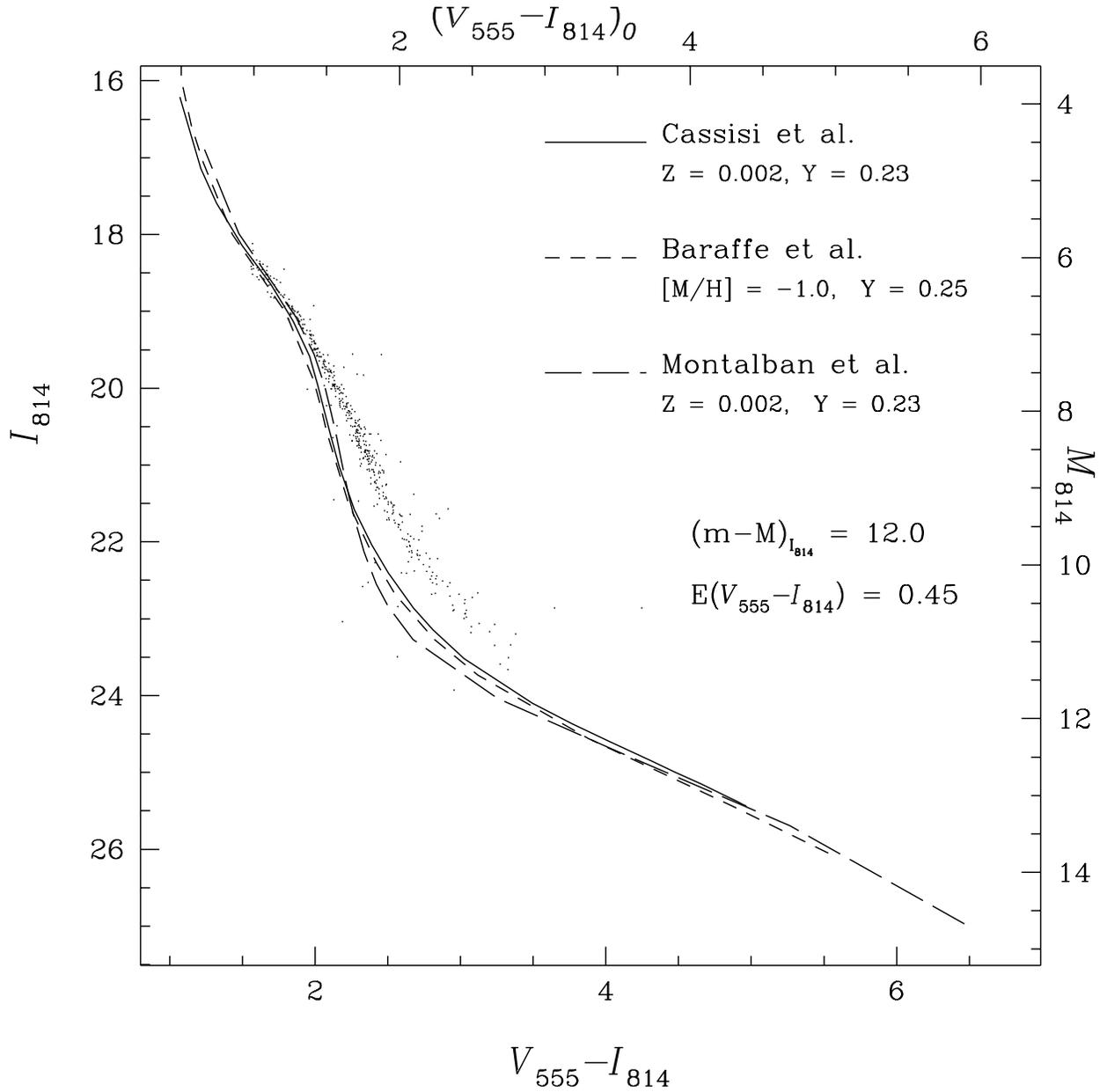}
\caption{ Comparison of theoretical colors and magnitudes with the 
observed sample of M4 main-sequence stars.
\label{fit} }
\end{figure}
\clearpage 

\begin{figure}
\plotone{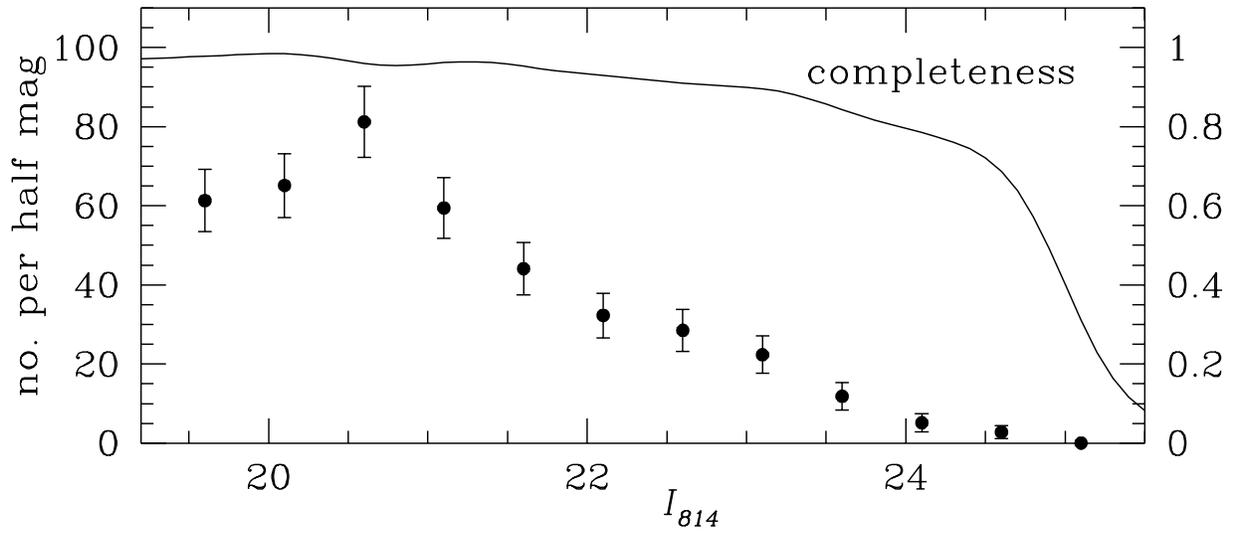}
\caption{Luminosity function of our field in M4.  Numbers have been
corrected for completeness, which is also shown separately.
Binning has been chosen so that the next-to-last bin ends at the point
of 50\% completeness.} 
\label{LFfig}
\end{figure}
\clearpage 

\begin{figure}
\plotone{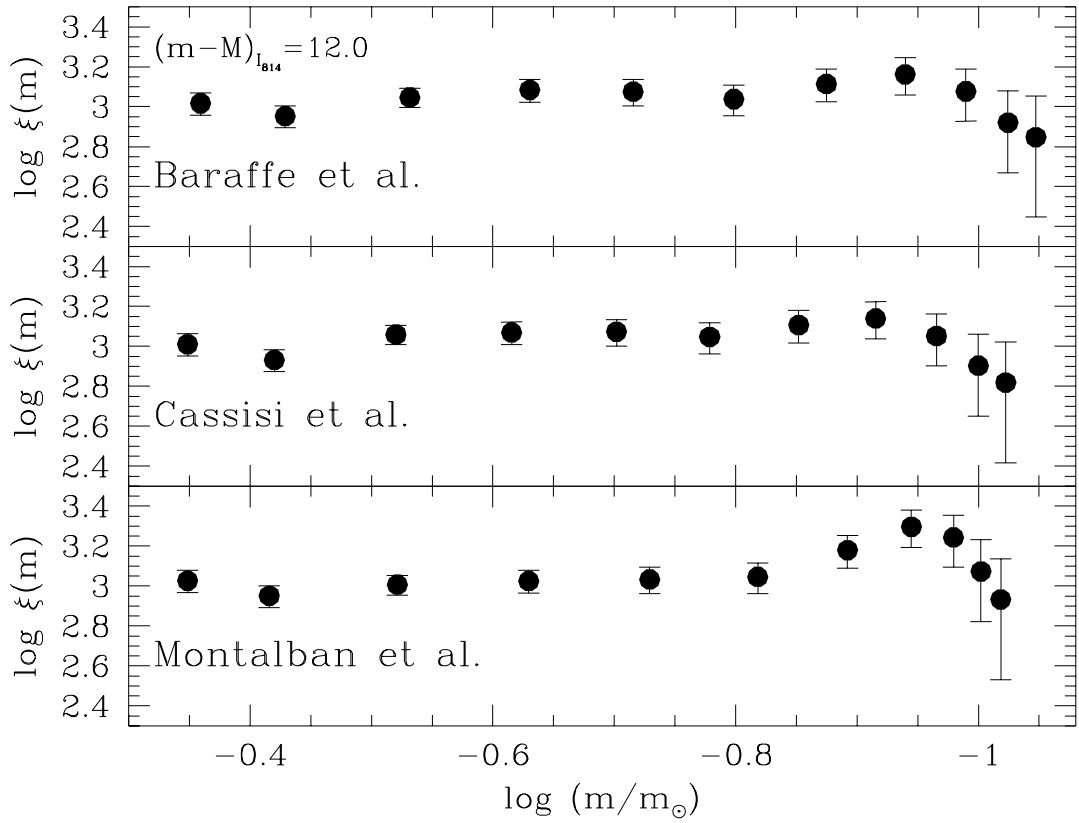}
\caption{Mass functions
(number of stars per unit mass) for our M4 field, calculated from MLRs
from three theoretical groups.}
\label{MFs}
\end{figure}

\end{document}